\documentclass[aps,prb,superscriptaddress,twocolumn,floatfix]{revtex4-2}
\usepackage[abs]{overpic}
\usepackage{graphicx,graphics}
\usepackage{dcolumn}
\usepackage{amsmath,amssymb,amsfonts}
\usepackage{wasysym}
\usepackage{latexsym,verbatim}
\usepackage{bm}
\usepackage{color}
\usepackage[framemethod=default]{mdframed}
\usepackage[breaklinks=true,colorlinks,citecolor=blue,linkcolor=blue,urlcolor=blue]{hyperref}
\usepackage[makeroom]{cancel}
\usepackage{fancybox}
\usepackage{ulem}

\newcommand{\be}{\begin{eqnarray}}
\newcommand{\ee}{\end{eqnarray}}

\begin{document}


\title{Ordered phases and superconductivity in two-dimensional electron systems subject to pair spin-orbit interaction}

\author{Feng Liu}
\email{feng.liu-6@postgrad.manchester.ac.uk}
\author{Alessandro Principi}%
\email{alessandro.principi@manchester.ac.uk}
\affiliation{%
Department of Physics and Astronomy, University of Manchester, M13 9PL Manchester (UK)
 }%


\begin{abstract}
Pair spin-orbit interaction can emerge in strongly-interacting systems characterized by a large spin-orbit coupling. Here we study the role of this interaction in stabilizing ordered and unconventional superconducting phases. We find that, if the system avoids superconductivity, the order realized is a combination of charge-density and spin-vorticity waves. The latter is reminiscent of a loop-current state, albeit in the spin, rather in the charge, channel. If the system becomes superconducting, the order parameter assumes the form of a paired density wave, {\it i.e.} pairing occurs at finite momentum. Intriguingly, one of the possible pairings acquires a form analogous to Amperean superconductivity. However, the order parameter here is always a blend of paired density wave and Amperean pairing, rather than being purely one or the other.
\end{abstract}

\maketitle


\section{INTRODUCTION}
The spin-orbit interaction (SOI), which couples the spin degree of freedom of electrons to their orbital motion, has received increasing attention in recent years~\cite{gindikin2020pair,sinova2004universal,li2015optical,mawrie2016drude,maiti2015collective}. Several spintronics applications rely in fact on strong SOI to give rise to novel effects that can be employed in functional devices.
More recently, the interplay between the SOI in Rashba materials and the Coulomb repulsive interaction has been shown to gives rise to a plethora of unusual physical effects~\cite{chen1999exchange,agarwal2011plasmon,xing2013spin,sun2005quantum,chesi2011high,liu2020phase}. For instance, it can lead to enhanced Zeeman splitting effects and modifications in the plasmon mass, as well as to the suppression of the Drude weight in certain systems.
In two-dimensional systems, the Rashba SOI is typically induced by an electric field perpendicular to the system which is the result of broken symmetry due to, e.g., the material's layered nature~\cite{bihlmayer2015focus,winkler2003spin,bihlmayer2022rashba,eremeev2012ideal}. In addition to this electric field, which responsible for the one-body SOI,
electron-electron interactions and the {\it pairwise} electric field they generate can also give rise to a SOI, known as {\it pair} spin-orbit interaction (PSOI). 

Although traditionally considered to be a small effect, recent studies have indicated that the PSOI is significantly enhanced in Rashba materials \cite{winkler2003spin,gindikin2020pair,gindikin2018spin}, leading to substantial modifications of non-interacting electronic states~\cite{bethe2012quantum,winkler2003spin,gindikin2020pair}. The diverse range of novel phenomena arising from this interaction is primarily attributed to the effective attraction that it can introduce amongst electrons~\cite{gindikin2022electron,gindikin2020pair}. Furthermore, unlike the Coulomb interaction, dimensional analysis reveals that the strength of PSOI interaction increases as the dimensionless Coulomb coupling parameter ${r_s=1/(a_B\sqrt{\pi n})}$ decreases. This implies that PSOI correlations can become dominant at sufficiently high densities~\cite{gindikin2022electron}. This contrasts with the conventional phenomenology of electron liquids, whereby interactions become dominant only in the low-density limit~\cite{giuliani2005quantum}.

Recently, it has been shown that for a PSOI two-dimensional electron system~\cite{gindikin2022electron}, once the particle density reaches a critical value, a sharp peak appears in the static structure factor. The peak is at a finite wave vector $q_c$. The density-density correlation function also exhibits a divergence at the same value of the wave vector~\cite{gindikin2022electron}. These signatures are typically due to a phase transition associated with the formation of a charge density wave (CDW)~\cite{kim1999new}. However, CDWs are known to be suppressed by the presence of long-range Coulomb interactions. 
This fact raises a question about the form of the order parameter in two-dimensional PSOI systems.

In this paper we investigate the type of order parameter associated with the phase transition observed in PSOI systems~\cite{kopietz2008bosonization,sengupta2005mott}.
We employ a double Hubbard-Stratonovich (H-S) transformation to derive an effective theory for a PSOI two-dimensional electron systems near the phase transition point. Using a mean-field approximation, and assuming that superconductivity does not occur, we derive an effective action for two competing order parameters, CDW and spin-current-vorticity wave (SCVW). There are averages of macroscopic observables which naturally emerge from the decoupling of Coulomb interaction and PSOI. While the CDW is characterised by a periodic modulation of the charge density, the unusual SCVW state exhibits periodic variations of the vorticity of the spin current. Since in the latter state spin-up currents locally circulate in directions opposite to spin-down ones, time reversal symmetry is not broken.

Near the phase transition point, the so-derived effective action for order parameters can be expressed in terms of correlation functions of these macroscopic quantities evaluated in the normal state. By finding the zero mode of the matrix coupling these macroscopic quantities, we determine the form of the order parameter driving the phase transition. We find that this is a linear combination of CDW and SCVW. Therefore, when the electron density is high enough, the attractive coupling due to the PSOI competes with the conventional long-range Coulomb interaction and can stabilise a CDW in a 2D electron gas.
As expected, the correlation function of the order parameter diverges at the phase transition point and at the wave vector of the CDW/SCVW modulation within the mean-field approximation. This divergence reflects simultaneously in both the density and spin-vorticity correlation functions, thus explaining the observations of Ref.~\cite{gindikin2022electron}. 
 
On the other hand, due to the effective attractive potential introduced by PSOI interactions, the system can also exhibit an instability towards a paired density wave (PDW) state~\cite{chen2004pair,soto2014pair,schmitt1989pairing}. We investigate this possibility by considering local (delta-function-like) Coulomb interactions and PSOI. We find that the PSOI favours pairing at finite momentum, among which an Amperean-like superconducting coupling between spin-currents. 

The paper is organized as following. In Sect.~\ref{sect:model} we introduce the model of electrons interacting via PSOI. In Sec.~\ref{Hs}, we derive the effective action for this Hamiltonian base on a double H-S transformation. In Sec.~\ref{Me}, we compute the free energy and correlation functions of the system using a mean-field approximation, and show the form acquired by the order parameter. The discussion regarding PDW state is presented in the Sec.~\ref{BCS}. Finally, we give a summary of results in Sec.~\ref{Con}.

\section{The model}
\label{sect:model}
We consider the following model Hamiltonian for a 2D electron gas interacting via the usual non-retarded Coulomb interaction and via a PSOI coupling, first introduced in Ref.~\cite{gindikin2022electron}
\begin{widetext}
\begin{align} \label{eq:Hamiltonian}
H=&-\frac{1}{2m}\sum_s\int d\mathbf{r}\psi^{\dagger}_s(\mathbf{r})\nabla^2_\mathbf{r}\psi_s(\mathbf{r})+\frac{1}{2}\sum_{s_1s_2}\int d\mathbf{r}_1d\mathbf{r}_2
\psi^{\dagger}_{s_1}(\mathbf{r}_1)\psi^{\dagger}_{s_2}(\mathbf{r}_2)\mathcal{U}(\mathbf{r}_1-\mathbf{r}_2)\psi_{s_2}(\mathbf{r}_2)\psi_{s_1}(\mathbf{r}_1)\nonumber
\\
&-i\alpha\sum_{s_1s_2}\int d\mathbf{r}_1d\mathbf{r}_2
\psi^{\dagger}_{s_1}(\mathbf{r}_1)\psi^{\dagger}_{s_2}(\mathbf{r}_2)[(s_1\partial_{x_1}-s_2\partial_{x_2})\mathcal{E}_y(\mathbf{r}_1-\mathbf{r}_2)-(s_1\partial_{y_1}-s_2\partial_{y_2})\mathcal{E}_x(\mathbf{r}_1-\mathbf{r}_2)]\psi_{s_2}(\mathbf{r}_2)\psi_{s_1}(\mathbf{r}_1)
\end{align}
\end{widetext}
Here $\psi_{s}(\mathbf{r})$ is the electron field operator, $s=\pm 1$ is the electron spin, $\mathcal{U}(\mathbf{r})=e^2/r$ is the Coulomb interaction potential, $\mathcal{E}(\mathbf{r})=e^{-1}\nabla\mathcal{U}(\mathbf{r})$ is the Coulomb field that produces the PSOI, and $\alpha$ stands for Rashba constant of the material. By introducing dimensionless parameters, {\it i.e.} by re-scaling real-space vectors with $\mathbf{r}\equiv\mathbf{r}/(r_sa_B)$ and introducing the dimensionless Rashba constant $\alpha\equiv\alpha/ea_B^2$, we can re-express the Coulomb interaction and field in the following form~\cite{fetter2012quantum,gindikin2022electron}
\begin{eqnarray}
    \mathcal{U}_{\mathbf{q}}&\rightarrow& r_s\mathcal{U}_{\mathbf{q}}
    \\
    \alpha\mathcal{E}(\mathbf{q})&\rightarrow&\frac{\alpha}{r_s}\mathcal{E}(\mathbf{q})
\end{eqnarray}
Therefore, in the limit of large densities, $r_s \to 0$, the PSOI becomes the dominant interaction in the Hamiltonian~(\ref{eq:Hamiltonian}).

\section{EFFECTIVE ACTION NEAR THE PHASE TRANSITION POINT}
\label{Hs}
In this section we perform the double H-S transformation that allows us to decouple the interaction terms in Eq.~(\ref{eq:Hamiltonian}) and bring it to a form amenable to mean-field approximations. To do so, we write the partition function $Z$ as a functional integral over a Grassmann field $\phi_s(\mathbf{r},\tau)$ with action 
%
\begin{align} \label{eq:action_phi_def}
    S[\phi^*,\phi]
    =\sum_s \int_0^{\beta}d\tau\int d\mathbf{r}\big[\phi^*_s(\mathbf{r},\tau)\partial_{\tau}\phi_s(\mathbf{r},\tau)+H\big]
\end{align}
where $H$ here is obtained from Eq.~(\ref{eq:Hamiltonian}) by replacing the operators $\psi_s(\mathbf{r})$ [$\psi^{\dagger}_s(\mathbf{r})$] with the Grassmann fields $\phi_s(\mathbf{r},\tau)$ [$\phi^*_s(\mathbf{r},\tau)$], $\tau$ is the imaginary time and $\beta$ is the inverse temperature. We call $S_{\text{int}}[\phi^*,\phi]$ the part of the action stemming from the interaction terms of Eq.~(\ref{eq:Hamiltonian}), and we rewrite it as
\begin{align} \label{eq:S_int_def}
    S_{\text{int}}[\phi^*,\phi]&=\frac{1}{2}\int_0^{\beta} d\tau d\tau' \int d\mathbf{r}d\mathbf{r'}\mathcal{U}(\mathbf{r}-\mathbf{r'})\delta_{\tau,\tau'}\nonumber
    \\
    &
   \times
    \big[ n(\mathbf{r},\tau)n(\mathbf{r'},\tau')+8m\alpha n(\mathbf{r},\tau)M(\mathbf{r'},\tau')\big]\nonumber
    \\
    &\equiv\frac{1}{2}(A\vert \mathcal{U}\vert A)-\frac{1}{2}(B\vert  \mathcal{U}\vert B),
\end{align}
%
{red}here we have introduced two functions corresponding to macroscopic physical quantities. The first $n(\mathbf{r},\tau)$
\begin{equation} \label{eq:density_def}
n(\mathbf{r},\tau) = \sum_s \phi^*_s(\mathbf{r},\tau)\phi_s(\mathbf{r},\tau)
\end{equation}
%
{red}represents the electron density, while the second $M(\mathbf{r},\tau)$ is expressed as:
\begin{align} \label{eq:M_def}
M(\mathbf{r},\tau)&=\frac{1}{2}\sum_{s}s\nabla_{\mathbf{r}}\times \mathbf{j}_{s}(\mathbf{r},\tau)=-\frac{1}{2}\sum_ss\nabla^2 \tilde{M}_s(\mathbf{r},\tau).
\end{align}
%
we refer to it as the spin-current-vorticity density. In Eq.~(\ref{eq:M_def}), $j_s(\mathbf{r},\tau)$ and $\tilde{M}_s(\mathbf{r},\tau)$ are the current density and magnetization for a given spin projection, respectively, {\it i.e.}
\begin{align}
j_s(\mathbf{r},\tau) &= \frac{1}{2mi}[\phi^*_s(\mathbf{r},\tau)\nabla\phi_s(\mathbf{r},\tau)-(\nabla \phi^*_s(\mathbf{r},\tau))\phi_s(\mathbf{r},\tau)]\nonumber
\\
&=\nabla\times \tilde{M}_s(\mathbf{r},\tau)
\end{align}
%
In Eq.~(\ref{eq:S_int_def}) we also introduced the notation
\begin{equation} \label{eq:scalar_prod1}
(\alpha \vert \mathcal{O}\vert \beta) = 
\int_0^{\beta} d\tau d\tau' \int d\mathbf{r}d\mathbf{r'}\mathcal{O}(\mathbf{r}-\mathbf{r'},\tau-\tau') \alpha(\mathbf{r},\tau) \beta(\mathbf{r'},\tau')
\end{equation}
with $\mathcal{U} \equiv \mathcal{U}(\mathbf{r}-\mathbf{r'})\delta_{\tau,\tau'}$, and $\alpha(\mathbf{r},\tau)$ and $\beta(\mathbf{r},\tau)$ are two real variables. Finally, In Eq.~(\ref{eq:S_int_def}) we denoted
\begin{align}
A(\mathbf{r},\tau)&=n(\mathbf{r},\tau)+4m\alpha M(\mathbf{r},\tau),
\nonumber
\\
B(\mathbf{r},\tau)&=4m\alpha M(\mathbf{r},\tau).
\end{align}
%
To decouple both fourth-order interaction terms simultaneously, we introduce two real auxiliary fields $\kappa$ and $\lambda$. By means of a H-S transformation, the partition function becomes~\cite{kopietz2008bosonization,stoof2009ultracold}
\begin{align} \label{eq:Z_HS1}
Z&=N^{-1}\int D[\phi^*,\phi;\kappa,\lambda]
\exp\bigg\{
-S_0[\phi^*,\phi]+\frac{1}{2}(\kappa\vert U^{-1}\vert\kappa)
\nonumber\\
&-\frac{1}{2}(\lambda\vert U^{-1}\vert \lambda)-(\kappa\vert A)-(\lambda\vert B)
\bigg\}
\nonumber
\\
&=N^{-1}\int D[\kappa,\lambda]e^{\frac{1}{2}(\kappa\vert U^{-1}\vert \kappa)-\frac{1}{2}(\lambda\vert U^{-1}\vert \lambda)-S_{\text{kin}}[\kappa,\lambda]},
\end{align}
where now
\begin{equation}
(\alpha \vert \beta) = 
\int_0^{\beta} d\tau \int d\mathbf{r} \alpha(\mathbf{r},\tau) \beta(\mathbf{r},\tau),
\end{equation}
which is obtained from Eq.~(\ref{eq:scalar_prod1}) when $\mathcal{O}(\mathbf{r}-\mathbf{r'},\tau-\tau') = \delta(\mathbf{r}-\mathbf{r'})\delta(\tau-\tau')$.
The normalization coefficient in Eq.~(\ref{eq:Z_HS1}) is defined as:
\begin{equation}
    N=\int D[\kappa,\lambda]\exp\left[\frac{1}{2}(\kappa\vert U^{-1}\vert \kappa)-\frac{1}{2}(\lambda\vert U^{-1}\vert \lambda)
    \right]
\end{equation}
%
To be able to conveniently compute quantities expressed in terms of collective density fluctuations, such as the density-density correlation function and the bosonized Hamiltonian, we now perform a second H-S transformation~\cite{kopietz2008bosonization}. We introduce two new real auxiliary fields $\xi$ and $\rho$, which can then be identified physically with the bosonized fluctuation (CDW and SCVW, respectively), to eliminate the fields $\kappa$ and $\lambda$.  Performing the integration over the fermionic fields ($\phi$ and $\phi^*$), the partition function becomes 
\begin{align} \label{eq:Z_secondHS}
Z&=\frac{N}{N}\int D[\rho,\xi]\exp\bigg[-\frac{1}{2}(\rho\vert U\vert\rho)+\frac{1}{2}(\xi\vert U\vert\xi)\bigg] \nonumber
\\
&\quad\times\int D[\kappa,\lambda]\exp\bigg\{ (\rho\vert\kappa)+(\xi\vert\lambda)-S_{\text{kin}}[\kappa,\lambda] \bigg\} \nonumber
\\
&=\int D[\rho,\xi]\exp\bigg\{-\frac{1}{2}(\rho\vert U\vert\rho)+\frac{1}{2}(\xi\vert U\vert\xi)-{\tilde S}_{\text{kin}}[\rho,\xi] \bigg\}\nonumber
\\
&\equiv \int D[\rho,\xi]\exp\left\{-\tilde{S}[\rho,\xi]
\right\}
\end{align}
where the kinetic part of the action $S_{\text{kin}}[\kappa,\lambda]$ is defined as
\begin{equation}
S_{\text{kin}}[\kappa,\lambda] = -\text{Tr}\ln(1-G_0\Sigma)
\end{equation}
with
\begin{align} \label{eq:Sigma_def}
   &\Sigma_{ss'}(\mathbf{r},\tau;\mathbf{r'},\tau')=\delta_{ss'}[\kappa(\mathbf{r},\tau)\delta(\mathbf{r}-\mathbf{r'})\delta(\tau-\tau')\nonumber
   \\
   &-i2\alpha s(\nabla_{\mathbf{r}}(\kappa(\mathbf{r},\tau)-i\lambda(\mathbf{r},\tau)))\times(\nabla_{\mathbf{r'}}\delta(\mathbf{r'}-\mathbf{r}))\delta(\tau-\tau')]
\end{align}
This complicated expression, up to a constant, can be written in terms of a series as $S_{\text{kin}}[\kappa,\lambda]=\sum_{\ell=1}^{\infty}\ell^{-1}\text{Tr}[(G_0\Sigma)^\ell]$.
To obtain the last line of Eq.~(\ref{eq:Z_secondHS}), we performed the integration over the fields $\kappa$ and $\lambda$, and we introduced ${\tilde S}_{\text{kin}}[\rho,\xi]$ as the functional Fourier transformation of  $S_{\text{kin}}[\kappa,\lambda]$, {\it i.e.}
\begin{equation}
    e^{-{\tilde S}_{\text{kin}}[\rho,\xi]}=\int D[\kappa,\lambda]e^{
   (\rho\vert \kappa)+(\xi\vert \lambda)-S_{\text{kin}}[\kappa,\lambda]}.
    \label{eq:FunFou}
\end{equation}

Finally, Using the generating functional approach \cite{stoof2009ultracold}, The vacuum expectation values and correlation functions of H-S fields in Eqs.~(\ref{eq:density_def})-(\ref{eq:M_def}) can be connected to the expectation values of the fields $\rho$ and $\xi$ and their correlation functions as
\begin{align}
    &\langle n(\mathbf{r},\tau)\rangle=\langle \rho(\mathbf{r},\tau)\rangle_{\tilde{S}}-\langle \xi(\mathbf{r},\tau)\rangle_{\tilde{S}}\nonumber
    \\
    &\langle M(\mathbf{r},\tau)\rangle=(4m\alpha)^{-1}\langle \xi(\mathbf{r},\tau)\rangle_{\tilde{S}}
    \nonumber
    \\
    &\langle n(\mathbf{r}_1,\tau_1)n(\mathbf{r}_2,\tau_2)\rangle\nonumber
    \\
    &\qquad=\langle (\rho(\mathbf{r}_1,\tau_1)-\xi(\mathbf{r}_1,\tau_1))(\rho(\mathbf{r}_2,\tau_2)-\xi(\mathbf{r}_2,\tau_2))\rangle_{\tilde{S}}\nonumber
    \\
    &\langle M(\mathbf{r}_1,\tau_1)M(\mathbf{r}_2,\tau_2)\rangle=(4m\alpha)^{-2}\langle \xi(\mathbf{r}_1,\tau_1)\xi(\mathbf{r}_2,\tau_2)\rangle_{\tilde{S}}
    \label{eq:mean}
\end{align}
Note, however, that the averages on the left- and right-hand side of these equations are taken with the actions of Eqs.~(\ref{eq:action_phi_def}) and~(\ref{eq:Z_secondHS}), respectively. Specifically, the two expressions can be written as follows:
\begin{eqnarray}
\langle O(\mathbf{r},\tau)\rangle &=&\frac{1}{Z}\int D[\phi^*,\phi]O(\mathbf{r},\tau)e^{-S[\phi^*,\phi]}
\\
\langle O(\mathbf{r},\tau)\rangle_{\tilde{S}}&=& \frac{1}{Z}\int D[\rho,\xi]O(\mathbf{r},\tau)e^{-\tilde{S}[\rho,\xi]}
\end{eqnarray}

Thus, by employing two consecutive H-S transformations
we have obtained
an effective action involving real fields, which is amenable to further mean-field treatment 
in the vicinity of the phase transition point, as we now proceed to show. We have also shown that the averages of the physical observables $n(\mathbf{r},\tau)$ and $M(\mathbf{r},\tau)$ can be obtained from the averages of the new H-S fields we have introduced.

\section{MEAN-FIELD APPROXIMATION}\label{Me}
The mapping shown in the previous section is exact, and therefore the final action in Eq.~(\ref{eq:Z_secondHS}) still exhibits all the complexity of the initial problem. To proceed further, it is necessary to approximate it. The simplest, but at the same time most powerful approximation enabling further analytical progress is a mean-field one. We now describe how such approximation is constructed.

We begin by truncating $S_{\text{kin}}[\kappa,\lambda]$ to quadratic order in the H-S fields as
\begin{align}
    S_{\text{eff}}[\kappa,\lambda]\simeq\text{Tr}[G_0\Sigma]+\frac{1}{2}\text{Tr}[(G_0\Sigma)^2].
    \label{eq:gaussian}
\end{align}
We note that for any electrically neutral system, the Hartree contributions from the first term in~(\ref{eq:gaussian}) will always be canceled out by contributions arising from the positive charge background. Therefore, we neglect it in what follows and focus on the terms arising from the last one in Eq.~(\ref{eq:gaussian}). Substituting Eq.~(\ref{eq:gaussian}) into Eq.~(\ref{eq:FunFou}), after some lengthy but straightforward algebra we obtain the effective action under Gaussian approximation
\begin{eqnarray}
    \frac{\tilde{S}[\rho,\xi]}{V\beta}&=&\frac{1}{2}\sum_{q}\rho_{-q}[\mathcal{U}(\mathbf{q})+\Pi_0^{-1}(q)]\rho_q\nonumber
    \\
    &+&\frac{1}{2}\sum_{q}\xi_{-q}[-\mathcal{U}(\mathbf{q})+\Pi_0^{-1}(q)+\Pi_1^{-1}(q)]\xi_{q}\nonumber
    \\
    &-&\frac{1}{2}\rho_{-q}\Pi_0^{-1}(q)\xi_q-\frac{1}{2}\xi_{-q}\Pi_0^{-1}(q)\rho_q,
    \label{eq:gaussact}
\end{eqnarray}
where $q=(\mathbf{q},i\omega_n)$ and
\begin{eqnarray}
    \rho_q&=&\frac{1}{V\beta}\int_0^\beta d\tau\int d\mathbf{r}\rho(\mathbf{r},\tau)e^{-i\mathbf{q}\cdot\mathbf{r}+i\omega_n\tau}\nonumber
    \\
    \xi_q&=&\frac{1}{V\beta}\int_0^\beta d\tau\int d\mathbf{r}\xi(\mathbf{r},\tau)e^{-i\mathbf{q}\cdot\mathbf{r}+i\omega_n\tau}\nonumber
    \\
    G_0(q)&=&\frac{1}{i\omega_n-\epsilon_{\mathbf{k}}+\mu}\nonumber
    \\
    \Pi_0(q)&=&\frac{-1}{V\beta}\sum_{k;s}G_0(k)G_0(k+q)\nonumber
    \\
    \Pi_1(q)&=&\frac{-(2\alpha)^2}{V\beta}\sum_{k;s}G_0(k)G_0(k+q)(\mathbf{q}\times \mathbf{k})^2
\end{eqnarray}

The next step to derive a mean-field action consists in performing a saddle-point approximation. We replace $\rho_q$ and $\xi_{q}$ with their average values in Eq.~(\ref{eq:gaussact}) and, using Eq.~(\ref{eq:mean}), we replace such averages with $\langle n_{-\mathbf{q}}\rangle$ and $\langle M_{-\mathbf{q}}\rangle$. The mean-field Landau-Ginzburg free-energy density which includes both Coulomb interaction and PSOI reads
\begin{equation} \label{eq:MF_GL_freeenergy}
    f[\langle n\rangle,\langle M\rangle] =\frac{1}{2}\sum_{\mathbf{q}}\left(\begin{array}{c}\langle n_{-\mathbf{q}}\rangle 
    \\\langle M_{-\mathbf{q}}\rangle\end{array}\right)^T\Gamma(\mathbf{q})
    \left(\begin{array}{c}\langle n_{\mathbf{q}}\rangle 
    \\\langle M_{\mathbf{q}}\rangle\end{array}\right),
\end{equation}
where
\begin{equation} \label{eq:MF_GL_freeenergy_Gamma_def}
    \Gamma(\mathbf{q})=\left(\begin{matrix}\mathcal{U}(\mathbf{q})+\Pi_0^{-1}(\mathbf{q})&4\alpha\mathcal{U}(\mathbf{q})
    \\
   4\alpha\mathcal{U}(\mathbf{q})&(4\alpha)^2\Pi_1^{-1}(\mathbf{q})
    \end{matrix}\right).
\end{equation}
We observe that, since the matrix $\Gamma(\mathbf{q})$ is even with respect to $\mathbf{q}$,
its eigenvectors always satisfy the following identity
\begin{equation}
\left(\begin{array}{c}\langle n_{-\mathbf{q}}\rangle 
    \\\langle M_{-\mathbf{q}}\rangle\end{array}\right) =
    \left(\begin{array}{c}\langle n_{\mathbf{q}}\rangle 
    \\\langle M_{\mathbf{q}}\rangle\end{array}\right).
\end{equation}
We now analyze Eqs.~(\ref{eq:MF_GL_freeenergy})-(\ref{eq:MF_GL_freeenergy_Gamma_def}) in more detail around the phase transition point. We determine the form of the order parameter, which is non-uniform in space, and the wave vector at which its modulation occur.

\subsection{Landau-Ginzburg free energy and order parameter}
The system is unstable against developing order, e.g. a CDW or SCVW (or a combination of the two -- see below) when an eigenvalue of the matrix $\Gamma(\mathbf{q})$ in Eq.~(\ref{eq:MF_GL_freeenergy_Gamma_def}) becomes zero or negative. The eigenvalues and eigenvectors of $\Gamma(\mathbf{q})$ are found by applying a unitary transformation that introduces new bosonic fields $\gamma_{\pm}(\mathbf{q})$ \cite{giuliani2005quantum}, {\it i.e.}
\begin{eqnarray}
    \langle n_{\mathbf{q}}\rangle&=&\gamma_+(\mathbf{q})\cos\theta_{\mathbf{q}}-\gamma_{-}(\mathbf{q})\sin\theta_{\mathbf{q}},
    \\
    \langle M_{\mathbf{q}}\rangle&=&\gamma_+(\mathbf{q})\sin\theta_{\mathbf{q}}+\gamma_{-}(\mathbf{q})\cos\theta_{\mathbf{q}},
\end{eqnarray}
where the rotation angle $\theta_{\mathbf{q}}$ is chosen so as to cancel off-diagonal terms and satisfies
\begin{equation}
    \tan (2\theta_{\mathbf{q}})=\frac{2\Gamma_{12}(\mathbf{q})}{\Gamma_{11}(\mathbf{q})-\Gamma_{22}(\mathbf{q})}.
    \label{eq:tant} 
\end{equation}
Using this transformation, the free energy can be diagonalized as
\begin{equation}
f=\frac{1}{2}\sum_{\mathbf{q},\pm}\epsilon_{\pm}(\mathbf{q})\gamma_{\pm}(-\mathbf{q})\gamma_{\pm}(\mathbf{q}),
\end{equation}
where 
\begin{eqnarray} \label{eq:epsilon_pm}
    \epsilon_{\pm}(\mathbf{q})&=&\frac{\Gamma_{11}(\mathbf{q})+\Gamma_{22}(\mathbf{q})}{2}\nonumber
    \\
    &\pm&\sqrt{\left(\frac{\Gamma_{11}(\mathbf{q})-\Gamma_{22}(\mathbf{q})}{2}\right)^2+\left(\Gamma_{12}(\mathbf{q})\right)^2}.
\end{eqnarray}
Using the zero-temperature expressions for $\Pi_0(\mathbf{q})$ and $\Pi_1(\mathbf{q})$ given in Ref.~\cite{mihaila2011lindhard,gindikin2022electron}, as well as the two-dimensional Coulomb interaction $\mathcal{U}(\mathbf{q})=2\pi/\vert \mathbf{q}\vert$, we can determine the eigenvalues $\epsilon_-(\mathbf{q})$ and $\epsilon_{+}(\mathbf{q})$. 

\begin{figure}[t]
	\begin{overpic}[width=0.99\linewidth]
            {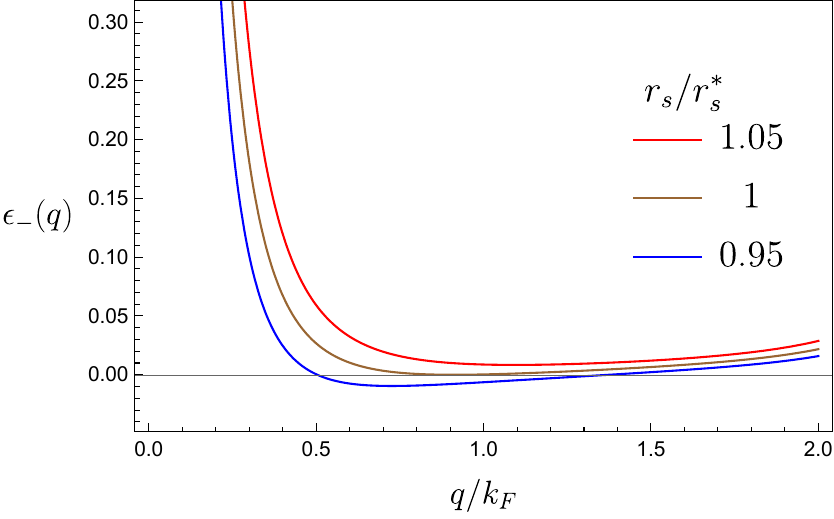}
            \put(42,130){\large{(a)}}
        \end{overpic}
	\begin{overpic}[width=0.97\linewidth]
            {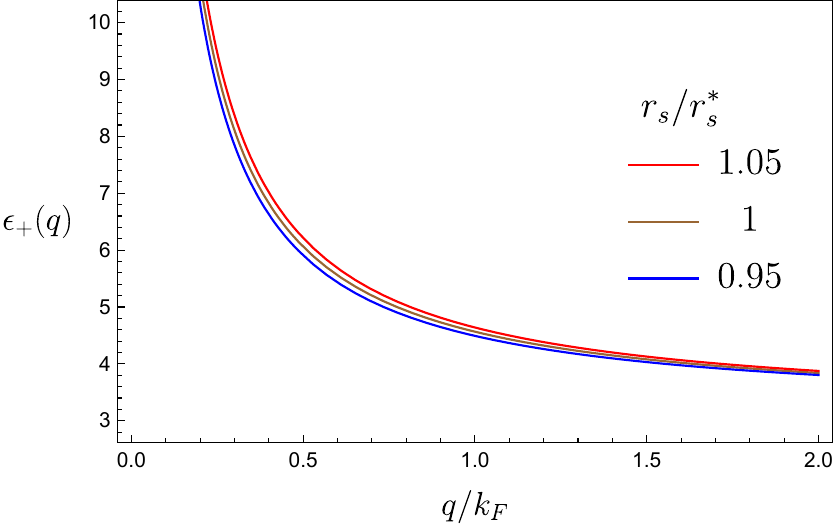}
            \put(40,130){\large{(b)}}
        \end{overpic}
\caption{\label{fig:fe1} 
Panel (a) the eigenvalue $\epsilon_-(q)$ in Eq.~(\ref{eq:epsilon_pm}) as a function of wave vector $q$ for a 2D system of non-relativistic electrons interacting via non-retarded Coulomb interaction and PSOI. The function exhibits a minimum that touches the horizontal axis for $r_s = r_s^*$ [given in Eq.~(\ref{eq:rs_star})]. For this value of $r_s$, the minimum is located at $q=q_c$, where $q_c$ is given in Eq.~(\ref{eq:qc}).
Panel (b) same as panel (a) but for the eigenvalue $\epsilon_+(q)$. The function is monotonically decreasing and it never vanishes.
In both plots we have set $\alpha=0.1$, $e=1$ and $m=1$,  which corresponds to $q_c\simeq0.91k_F$}.
\end{figure}

In Fig.~\ref{fig:fe1} we see that, while $\epsilon_+(\mathbf{q})$ is a monotonic function, $\epsilon_-(\mathbf{q})$ has a minimum at a finite wave vector whose magnitude depends on the value of $r_s$. At $r_s=r_s^*$ and $q = q_c$, the minimum value of $\epsilon_-(\mathbf{q})$ vanishes. The critical values of $r_s^*$ and $q_c$ are related to $\alpha$ according to~\cite{gindikin2022electron}
\begin{eqnarray}
    r_s^*&=&\frac{2^{\frac{13}{6}}\alpha}{\sqrt{2^{1/3}+2\alpha^{2/3}3^{2/3}}},
    \label{eq:rs_star}
    \\
     \frac{q_c}{k_F}&=&\frac{2\alpha^{1/3}3^{1/3}}{\sqrt{2^{1/3}+2\alpha^{2/3}3^{2/3}}}
     \label{eq:qc}.
\end{eqnarray}
These are depicted in Fig.~\ref{fig:fe1}(a) and (b), respectively, as a function of the parameter $r_s/r_s^*$. 
%
For these values of $r_s^*$ and $q_c$ the order parameter $\gamma_-(q_c)$ becomes nonzero and equal to
\begin{equation}
    \gamma_-(q_c)=-\sin\theta_{q_c}\langle n_{q_c}\rangle +\cos\theta_{q_c}\langle M_{q_c}\rangle.
\end{equation}

\begin{figure}[t]
\begin{overpic}[width=0.99\linewidth]
            {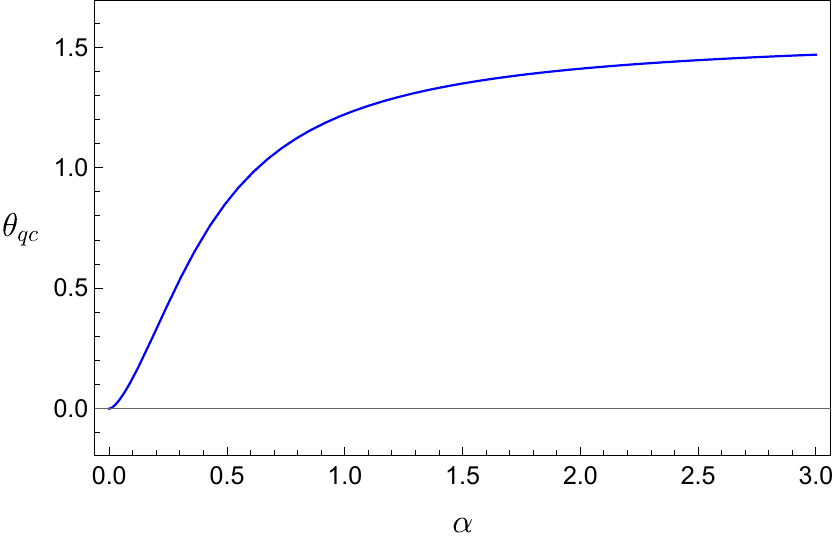}
        \end{overpic}
\caption{\label{fig:tht} 
The value of the mixing angle $\theta_{qc}$ between CDW and SCVW phases as a function of $\alpha$.  For small $\alpha$, the order parameter becomes a pure SCVW. Conversely, for large values of $\alpha$ it approaches a pure CDW. Therefore, the PSOI improves the stability of CDW broken symmetry phases compared to the conventional 2D electron gas.}
\end{figure}

By plugging the value of $q = q_c$ given in Eq.~(\ref{eq:qc}) into Eq.~(\ref{eq:tant}), we obtain the relationship between the angle $\theta_{q_c}$ and the PSOI parameter $\alpha$ at the critical value $r_s^*$, which is shown in Fig.~\ref{fig:tht}.
It is evident from Fig.~\ref{fig:tht} that, for any given parameter $\alpha$, the order parameter near the critical point always manifests as a mixture of CDW $\langle n_{q_c}\rangle$ and SCVW $\langle M_{q_c}\rangle$. When the parameter $\alpha$ approaches zero, also $\theta_{q_c}\rightarrow 0$. In this case, we can simplify the expression of the free energy and obtain
\begin{eqnarray}
f&=&\frac{1}{2}\sum_{\mathbf{q},\pm}\epsilon_{\pm}(\mathbf{q})\gamma_{\pm}(-\mathbf{q})\gamma_{\pm}(\mathbf{q})\nonumber
\\
&\rightarrow&
\frac{1}{2} \sum_{\vert\mathbf{q}\vert=q_c} \epsilon_{-}(\mathbf{q}_c)\gamma_{-}(-\mathbf{q}_c)\gamma_{-}(\mathbf{q}_c)\nonumber
\\
&\rightarrow&
\frac{1}{2} \sum_{\vert\mathbf{q}\vert=q_c} \epsilon_{-}(\mathbf{q}_c)\langle M_{-\mathbf{q}_c}\rangle\langle M_{\mathbf{q}_c}\rangle.
\end{eqnarray}
Therefore, the order parameter becomes a pure SCVW.
On the other hand, when the parameter $\alpha$ is very large, $\theta_{q_c}\rightarrow \pi/2$, and the effective free energy becomes
\begin{eqnarray}
    f&\rightarrow&
    \frac{1}{2} \sum_{\vert\mathbf{q}\vert=q_c} \epsilon_{-}(\mathbf{q}_c)\gamma_{-}(-\mathbf{q}_c)\gamma_{-}(\mathbf{q}_c) 
    \nonumber
    \\
    &\rightarrow&
    \frac{1}{2} \sum_{\vert\mathbf{q}\vert=q_c} \epsilon_{-}(\mathbf{q}_c)\langle n_{-\mathbf{q}_c}\rangle\langle n_{\mathbf{q}_c}\rangle,
\end{eqnarray}
{\it i.e.} the order parameter is a CDW.

To summarize, we found that the order parameter of a 2D electron gas in the presence of PSOI is always a linear combination of CDW and SCVW, rather than a single one of them. In the next subsection, we will also clarify that the corresponding static correlation functions [$\chi_{nn}(\mathbf{q}_c,0)$ and $\chi_{MM}(\mathbf{q}_c,0)$] are simultaneously divergent at the critical point, and therefore the phase transition can be detected in either channel. However, for small or large values of $\alpha$, the order parameter tends to a pure SCVW or CDW, respectively.
This result also implies that, for sufficiently weak PSOI interactions, the system is not expected to develop a CDW.
%
We stress that this does not mean that the system does not undergo a phase transition at all. 
Contrary to the case of the standard 2D electron gas, this mean-field result indicates that, at sufficiently high densities, the attraction due to the PSOI is always strong enough to counterbalance the Coulomb repulsion.
{However}, if the electron densities required become too high, the $\mathbf{k}\cdot\mathbf{p}$ approximation used to describe a real material's band structure with the Hamiltonian in Eq.~(\ref{eq:Hamiltonian}) may not be applicable. As a consequence, the results of this paper [based on Eq.~(\ref{eq:Hamiltonian})] { may} lose their validity~\cite{gindikin2022electron}. 
{Taking into account such band structure effects is beyond the scope of the present work.}
{Thus}, disregarding this aspect, it can be stated that in the mean-field approximation, a phase transition always occurs in a 2D electron gas at a sufficiently high density,
and it is always a mixture of SCVW and CDW.
It is also interesting to note that, for large values of the PSOI coupling $\alpha$, the order parameter acquires a significant CDW component, at odds with what expected in conventional 2D electron gases where a spin-density wave dominates ~\cite{bernu2011hartree}.

\subsection{Static correlation function}
\begin{figure}[t]
\begin{overpic}[width=0.97\linewidth]
            {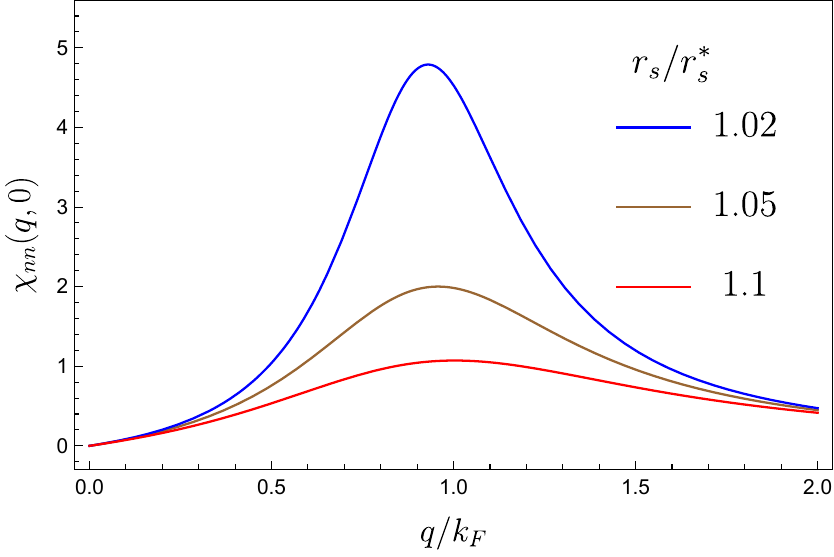}
            \put(30,135){\large{(a)}}
            \put(121,35){\color{black}\vector(0,-1){10}}
            \put(110,30){\color{black}$q_c$}
        \end{overpic}
\begin{overpic}[width=0.99\linewidth]
            {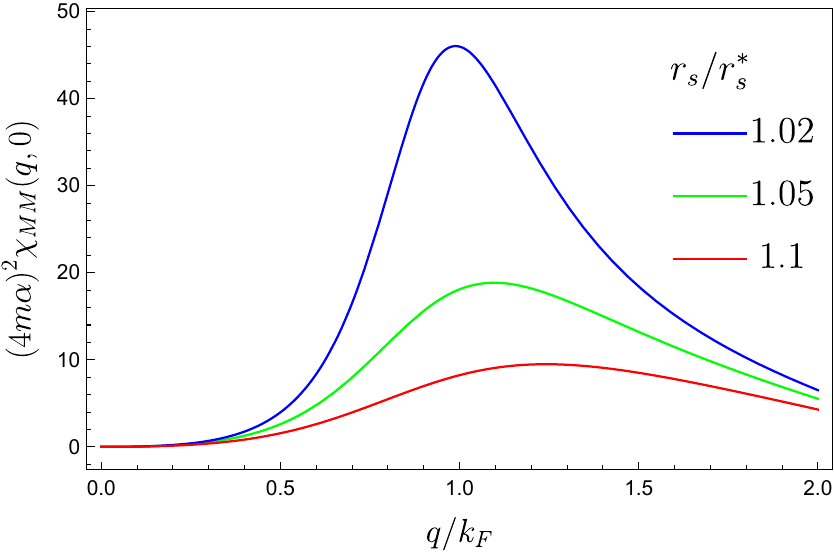}
            \put(30,135){\large{(b)}}
            \put(125,35){\color{black}\vector(0,-1){10}}
            \put(114,30){\color{black}$q_c$}
        \end{overpic}
\caption{\label{fig:cc1}
Panel (a) the density-density static correlation function plotted as a function of wave vector $q$ and for $\alpha=0.1$, $e=1$ and $m=1$ (which corresponds to $q_c\simeq0.91k_F$, as indicated by the black arrow in the figure).
Panel (b) the SCVW static correlation function plotted as a function of wave vector $q$.
Both functions shows a prominent peak which moves to the left and increases in magnitude when $r_s \to r_s^*$, signalling the onset of the phase transition at $r_s^\star$.}
\end{figure}

To derive the static correlation function, we now transform the effective action in Eq.~(\ref{eq:gaussact}) by introducing the fields $\zeta_\pm(q)$ as
\begin{eqnarray}
    \rho(q)&=&\zeta_-(q)\cos\vartheta_{q}-\zeta_+(q)\sin\vartheta_q,
    \\
    \xi(q)&=&\zeta_-(q)\sin\vartheta_q+\zeta_+(q)\cos\vartheta_q.
\end{eqnarray}
where $\vartheta$ here is another unitary transformation parameter introduced for the purpose of orthogonalizing with an action of  Eq.~(\ref{eq:gaussact}).
Using these relations, Eq.~(\ref{eq:gaussact}) becomes
\begin{equation}
    S[\zeta_-,\zeta_+]=\frac{V\beta}{2}\sum_{q}\zeta_{\pm}(-q)\Pi_{\pm}^{-1}(q)\zeta_{\pm}(q),
\end{equation}
where
\begin{eqnarray}
    \Pi_{\pm}^{-1}(q)&=&\Pi_{0}^{-1}(q)+\frac{1}{2}\Pi_{1}^{-1}(q) \nonumber
    \\
    &\pm&\sqrt{\left(\mathcal{U}(\mathbf{q})-\frac{\Pi_{1}^{-1}(q)}{2}\right)^2+\left(\Pi_{0}^{-1}(q)\right)^2}.
\end{eqnarray}
By utilizing the change of variables in Eq.~(\ref{eq:mean}) and setting $i\omega_n=0$, we can find the density-density and $M-M$ static correlation functions as
\begin{eqnarray}
    \chi_{nn}(\mathbf{q},0)&=&\cos^2\vartheta_{\mathbf{q}}\Pi_{-}(\mathbf{q})+\sin^2\vartheta_{\mathbf{q}}\Pi_{+}(\mathbf{q})\nonumber
    \\
    &-&2\cos\vartheta_{\mathbf{q}}\sin\vartheta_{q}(\Pi_{-}(\mathbf{q})-\Pi_{+}(\mathbf{q}))\nonumber
    \\
    &+&\sin^2\vartheta_{\mathbf{q}}\Pi_{-}(\mathbf{q})+\cos^2\vartheta_{\mathbf{q}}\Pi_{+}(\mathbf{q}),
    \\
    \chi_{MM}(\mathbf{q},0)&=&\frac{(\sin^2\vartheta_{\mathbf{q}}\Pi_{-}(\mathbf{q})+\cos^2\vartheta_{\mathbf{q}}\Pi_{+}(\mathbf{q}))}{(4m\alpha)^2}.
\end{eqnarray}
The two response functions are plotted in Fig.~\ref{fig:cc1}(a) and (b), respectively. Comparing the two figures, we see that they both exhibit peaks at the same values of $q$. The peaks shift to the left as the parameter $r_s/r_s^*$ approaches one, and grow in size. This behaviour clearly parallels that of $\epsilon_-(\mathbf{q})$, which vanishes for $q=q_c$ in the limit $r_s/r_s^* \to 1$. This fact shows that the phase transition due to PSOI, which occurs simultaneously in the density and SCVW channel, can be detected by measuring either correlation function.

\section{Pair-density wave state}\label{BCS}
The previous discussion has shown that, even in the presence of repulsive Coulomb interactions, the PSOI in a 2D electron system effectively provides an attractive potential and can induce phase transitions when the density is sufficiently high. We have studied phase transition in the CDW and SCVW channels.
In addition to the possibilities discussed above, here we analyze PDW and Amperean-like superconductivity. 
In both cases, Cooper pairs have non-zero momentum. 
Two-body calculations for 
PSOI-coupled electrons
indicate that 
bound states 
of 
particles of
equal spins are independent of the center-of-mass momentum~\cite{gindikin2018spin}. This makes it challenging to generate a PDW and Amperean-like superconductivity by condensing such pairs. {Therefore}, here we assume that only bound states formed with electrons of opposite spins ({also known as} convective bound states~\cite{gindikin2018spin}) participate in 
{the superconductivity}.

The discussion of the pairing instability here mirrors the derivations given in the previous sections.  Thus, we will be brief on several details, which are analogous to what discussed above.

We note that Amperean-like superconductivity can emerge thanks to the second term on the last line of Eq.~(\ref{eq:S_int_def}), which can be rewritten as
\begin{eqnarray} \label{eq:amperean_like_coupling}
    S_{\text{int}}^{(2)}&=&-\frac{1}{2}(B\vert  \mathcal{U}\vert B)\nonumber
    \\
    &=&-\frac{1}{2}\int_0^{\beta}d\tau\sum_{s_1s_2}s_1s_2\sum_{\mathbf{p},\mathbf{k},\mathbf{k'}}V_{\mathbf{k'}\mathbf{k}}(\mathbf{p})\phi^*_{s_1,\mathbf{k'}+\frac{\mathbf{p}}{2}}(\tau)
    \nonumber
    \\
    &\times&\phi_{s_2,-\mathbf{k'}+\frac{\mathbf{p}}{2}}^*(\tau)\phi_{s_2,-\mathbf{k}+\frac{\mathbf{p}}{2}}(\tau)\phi_{s_1,\mathbf{k}+\frac{\mathbf{p}}{2}}(\tau),
\end{eqnarray}
where the symmetrized interaction is
\begin{eqnarray}
    V_{\mathbf{k'}\mathbf{k}}(\mathbf{p})&=&4\alpha^2\mathcal{U}(\mathbf{k'}-\mathbf{k}) \left[(\mathbf{k'}-\mathbf{k})\times\frac{\mathbf{p}+\mathbf{k}+\mathbf{k'}}{2}\right]\nonumber
    \\
    &\times&\left[(\mathbf{k'}-\mathbf{k})\times\frac{\mathbf{p}-(\mathbf{k'}+\mathbf{k})}{2}\right].
\end{eqnarray}
%
The term in Eq.~(\ref{eq:amperean_like_coupling}) has a structure which is very similar to the conventional Amperean pairing~\cite{hugdal2020possible}. However, it describes an attractive interaction between {\it spins}, rather than charge currents, and therefore is not expected to produce a time-reversal-symmetry broken state.

We start by introducing the PDW and Amperean-like pairings 
\begin{eqnarray}
\Delta_{\mathbf{p}}(\mathbf{k},\tau)&\equiv&\Delta(-\mathbf{k}+\mathbf{p}/2,\mathbf{k}+\mathbf{p}/2,\tau)\nonumber
\\
&=&\phi_{\downarrow}(-\mathbf{k}+\mathbf{p}/2,\tau)\phi_{\uparrow}(\mathbf{k}+\mathbf{p}/2,\tau),
\\
\Lambda_{\mathbf{p}}(\mathbf{k},\tau)
&\equiv&\Lambda(-\mathbf{k}+\mathbf{p}/2,\mathbf{k}+\mathbf{p}/2,\tau)\nonumber
\\
&=&i2\alpha(\mathbf{k}\times \mathbf{p})\nonumber
\\
&\times&\phi_{\downarrow}(-\mathbf{k}+\mathbf{p}/2,\tau)\phi_{\uparrow}(\mathbf{k}+\mathbf{p}/2,\tau),
\end{eqnarray}
respectively,
which allow us to express the effective interaction action as
\begin{eqnarray}
S_{\text{int}}^{\text{eff}}[\phi^*,\phi]&=&\frac{1}{V}\int_0^{\beta}d\tau\sum_{\mathbf{p},\mathbf{k},\mathbf{k'}}\mathcal{U}_{\mathbf{k'}-\mathbf{k}}[\Delta_{\mathbf{p}}^*(\mathbf{k'},\tau)\Delta_{\mathbf{p}}(\mathbf{k},\tau)\nonumber
\\
&+&\Lambda_{\mathbf{p}}^*(\mathbf{k'},\tau)\Delta_{\mathbf{p}}(\mathbf{k},\tau)+\Delta_{\mathbf{p}}^*(\mathbf{k'},\tau)\Lambda_{\mathbf{p}}(\mathbf{k},\tau)]\nonumber
\\
&=&(\Delta\Vert \mathcal{U}\Vert\Delta)+(\Lambda\Vert \mathcal{U}\Vert\Delta)+(\Delta\Vert \mathcal{U}\Vert\Lambda)\nonumber
\\
&\equiv& (\Delta+\Lambda\Vert \mathcal{U}\Vert\Delta+\Lambda)-(\Lambda\Vert \mathcal{U}\Vert\Lambda)\label{eq:PDW}
\end{eqnarray}
where we introduced the notation \cite{stoof2009ultracold}
\begin{eqnarray}
(\alpha\Vert \mathcal{O} \Vert \beta)&=&\int_0^{\beta}d\tau d\tau'\int d\mathbf{r}d\mathbf{r'}
\alpha^*(\mathbf{r},\tau,\mathbf{r'},\tau')\nonumber
\\
&\times&\mathcal{O}(\mathbf{r}-\mathbf{r'},\tau-\tau')\beta(\mathbf{r'},\tau',\mathbf{r},\tau)
\\
(\alpha\Vert \mathcal{O}^{-1} \Vert \beta)&=&\int_0^{\beta}d\tau d\tau'\int d\mathbf{r}d\mathbf{r'}
\alpha^*(\mathbf{r},\tau,\mathbf{r'},\tau')\nonumber
\\
&\times&\frac{1}{\mathcal{O}(\mathbf{r}-\mathbf{r'},\tau-\tau')}\beta(\mathbf{r'},\tau',\mathbf{r},\tau).
\end{eqnarray}

We decompose both fourth-order interaction terms by introducing two H-S complex bosonic fields $X$ and $Y$ and integrate out the fields $\phi$ and $\phi^*$. We obtain
\begin{eqnarray}
    Z &=&
    \int D[X,Y]e^{-S[X,Y]},
    \label{eq:BCS}
\end{eqnarray}
where
\begin{eqnarray}
    S[X,Y]&=&-(X\Vert \mathcal{U}^{-1}\Vert X)+(Y\Vert \mathcal{U}^{-1}\Vert Y)\nonumber
    \\
    &+&\sum_{l=1}^{\infty}\frac{1}{l}\text{Tr}[(\mathbf{G}_0\boldsymbol{\Sigma})^l].
    \label{eq:BCS_action}
\end{eqnarray}
Here, the non-interacting Green’s function ($\mathbf{G}_0$) and self-energy ($\boldsymbol{\Sigma}$) matrices are
\begin{align}
&\mathbf{G}_0(\mathbf{x},\tau;\mathbf{x'},\tau')=\left[\begin{matrix}
G_0(\mathbf{x},\tau;\mathbf{x'},\tau')&0
\\
0&-G_0(\mathbf{x'},\tau';\mathbf{x},\tau)
\end{matrix}\right],
\\
&\boldsymbol{\Sigma}(\mathbf{x},\tau;\mathbf{x'},\tau')
=\left[\begin{matrix}
0&\Sigma_{12}(\mathbf{x},\tau;\mathbf{x'},\tau')
\\
\Sigma_{21}(\mathbf{x},\tau;\mathbf{x'},\tau')&0
\end{matrix}\right],
\end{align}
respectively, with
\begin{align}
\Sigma_{12}&(\mathbf{x},\tau;\mathbf{x'},\tau')=\delta(\tau-\tau')[X(\mathbf{x'},\mathbf{x},\tau)\nonumber
\\
&+i2\alpha (\nabla_{\mathbf{x}}\times \nabla_{\mathbf{x'}})(X(\mathbf{x'},\mathbf{x},\tau)+Y(\mathbf{x'},\mathbf{x},\tau))],
\\
\Sigma_{21}&(\mathbf{x},\tau;\mathbf{x'},\tau')=\delta(\tau-\tau')[X^*(\mathbf{x'},\mathbf{x},\tau)\nonumber
\\
&+i2\alpha  (\nabla_{\mathbf{x}}\times \nabla_{\mathbf{x'}})(X^*(\mathbf{x'},\mathbf{x},\tau)+Y^*(\mathbf{x'},\mathbf{x},\tau))].
\end{align}

Similar to the derivation of Eq.~(\ref{eq:mean}), the relationship between the vacuum expectation values of the H-S fields and the pairing fields is given by
\begin{eqnarray}
\langle X(\mathbf{x},\mathbf{x'},\tau)\rangle&=&\mathcal{U}(\mathbf{x}-\mathbf{x'})\langle \Delta(\mathbf{x},\mathbf{x'},\tau)+\Lambda(\mathbf{x},\mathbf{x'},\tau)
\rangle,
\nonumber
\\
\langle Y(\mathbf{x},\mathbf{x'},\tau)\rangle &=&-\mathcal{U}(\mathbf{x}-\mathbf{x'})\langle \Lambda(\mathbf{x},\mathbf{x'},\tau)\rangle.
\label{eq:mean2}
\end{eqnarray}
We will assume that mean-fields acquire the form~\cite{hugdal2020possible}
\begin{eqnarray}
    \langle \Delta_{\mathbf{p}}(\mathbf{k},\tau)\rangle &\equiv&\delta_{\mathbf{p},\mathbf{Q}}\Delta_{\mathbf{Q}}(\mathbf{k}),\nonumber
    \\
    \langle \Lambda_{\mathbf{p}}(\mathbf{k},\tau)\rangle &\equiv&\delta_{\mathbf{p},\mathbf{Q}}\Lambda_{\mathbf{Q}}(\mathbf{k}),
    \label{eq:mean3}
\end{eqnarray}
{\it i.e.} cooper pairs have a finite momentum in both the PDW and Amperean channels.

Within the Gaussian approximation, we truncate again the effective action to quadratic order in the self-energy. By substituting Eq.~(\ref{eq:mean2}) and~(\ref{eq:mean3}) into Eq.~(\ref{eq:BCS}), after some straightforward algebra we obtain the Landau-Ginzburg free energy density
\begin{eqnarray} \label{eq:PDW_GL_1}
    f&=&\left(\begin{array}{c} \Delta^*_{\mathbf{Q}} \\
    \Lambda^*_{\mathbf{Q}}\end{array}\right)^T
    \Gamma(\mathbf{Q})\left(\begin{array}{c}\Delta_{\mathbf{Q}} \\
    \Lambda_{\mathbf{Q}}\end{array}\right),
\end{eqnarray}
where we denoted as $\Delta_{\mathbf{Q}}$ and $\Lambda_{\mathbf{Q}}$ the vectors with components $\Delta_{\mathbf{Q}}(\mathbf{k})$ and $\Lambda_{\mathbf{Q}}(\mathbf{k})$, respectively, and
\begin{eqnarray} \label{eq:PDW_GL_2}
    \Gamma(\mathbf{Q})&=&-\left(\begin{matrix}\mathcal{U}+\mathcal{U}(\Pi_2+\Pi_3)\mathcal{U}&\mathcal{U}+\mathcal{U}\Pi_2\mathcal{U}
    \\
    \mathcal{U}+\mathcal{U}\Pi_2\mathcal{U}&\mathcal{U}\Pi_2\mathcal{U}
    \end{matrix}\right).
\end{eqnarray}
In these equations, every product includes integration over the momentum ${\bm Q}$ with measure $1/V$. In~(\ref{eq:PDW_GL_2}), $\mathcal{U}$, $\Pi_2$ and $\Pi_3$ are matrices with elements 
\begin{eqnarray}
    \mathcal{U}_{\mathbf{k},\mathbf{k'}}&=&\mathcal{U}(\mathbf{k}-\mathbf{k'}),
    \nonumber
    \\
    \Pi_{2,\mathbf{k},\mathbf{k'}}&=&\delta_{\mathbf{k},\mathbf{k'}}\frac{V}{\beta}\sum_{\omega_n}G_0(\mathbf{k},i\omega_n)G_0(\mathbf{Q}-\mathbf{k},-i\omega_n),
    \nonumber
    \\
    \Pi_{3,\mathbf{k},\mathbf{k'}}&=&(4\alpha^2)\delta_{\mathbf{k},\mathbf{k'}}\frac{V}{\beta}\sum_{\omega_n}(\mathbf{k}\times\mathbf{Q})^2\nonumber
    \\
    &\otimes&G_0(\mathbf{k},i\omega_n)G_0(\mathbf{Q}-\mathbf{k},-i\omega_n).
\end{eqnarray}

\begin{figure}[t]
\begin{overpic}[width=0.98\linewidth]
            {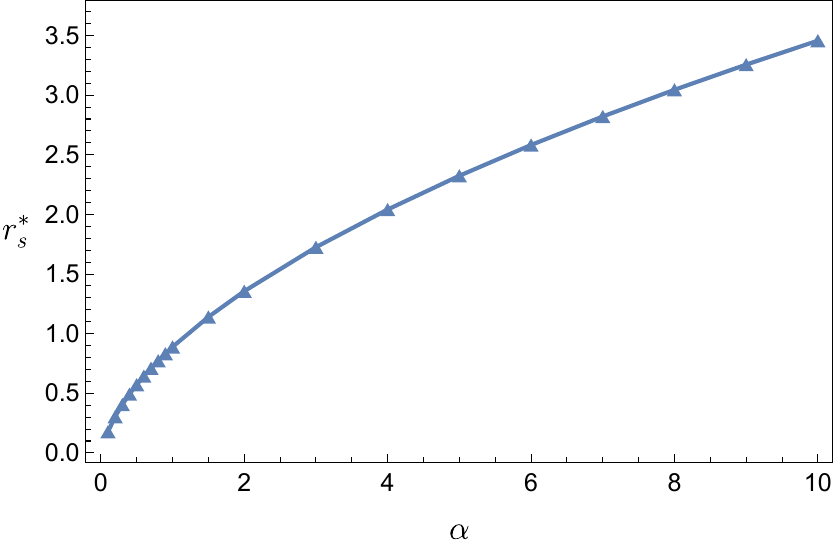}
            \put(35,130){\large{(a)}}
        \end{overpic}
\begin{overpic}[width=1.\linewidth]
            {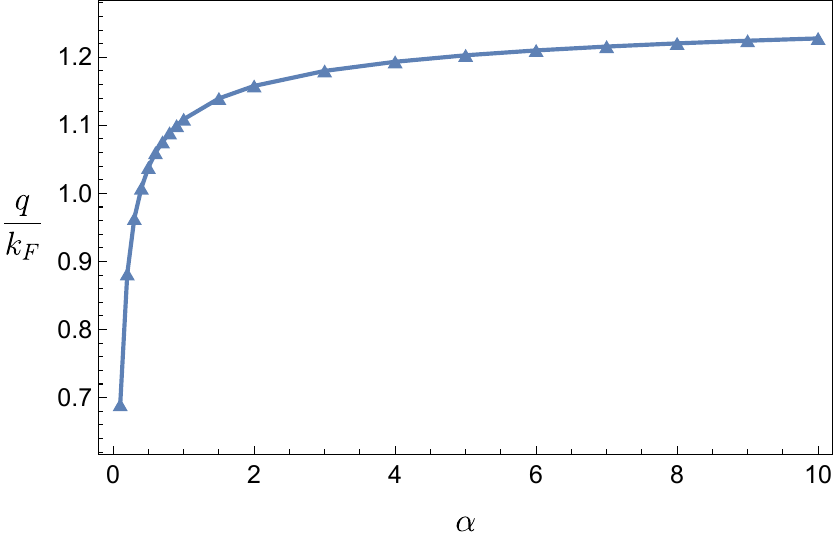}
            \put(35,130){\large{(b)}}
        \end{overpic}
\caption{\label{fig:br1}
Panel (a) the critical value of $r_s^*$ for the PDW state as a function of $\alpha$ for $U=1$ and $m=1$ with $\alpha$ ranging from $0.1$ to $10$.
Panel (b) the critical value $q_c$ for the PDW state as a function of $\alpha$. Other parameters are the same as in Panel (a). This mean-field result indicates that for any given $\alpha$, it it possible to find an electron density such that a PDW phase transition occurs. The value of $q_c$ corresponding to such phase transition increases as $\alpha$ increases.}
\end{figure}

In analogy to what shown in the previous sections, the system becomes unstable against developing superconducting phases when an eigenvalue of the kernel $\Gamma(\mathbf{Q})$ becomes zero~\cite{hugdal2020possible,lee2007amperean}. 
It is important to note that since $\vert \mathbf{Q}\vert \ne 0$, $\Delta_{\mathbf{Q}}(\mathbf{k})$ is generally not even for $\mathbf{k} \to -\mathbf{k}$. This implies that electron pairs do not only occur in the singlet channel but also possess a singlet-triplet structure (although the spins of the two electrons are always opposite), depending on the angle between $\mathbf{k}$ and $\mathbf{Q}$. This observation is consistent with previous two-body calculations~\cite{gindikin2018spin}.
For the general case of $\mathcal{U}(\mathbf{q})$, this eigenvalue problem is formidably complex and solving it requires employing heavily numerical methods. To gain qualitative insight in the instability, in this paper, we consider the case of a local (delta) interaction, whereby $\mathcal{U}(\mathbf{q}) = U$ for all values of $\mathbf{q}$. This greatly simplifies the Landau-Ginzburg free energy density, that yields
\begin{eqnarray}
    f&=&\left(\begin{array}{c} \tilde{\Delta}^*_{\mathbf{Q}} \\
    \tilde{\Lambda}^*_{\mathbf{Q}}\end{array}\right)^T
    \tilde{\Gamma}(\mathbf{Q})\left(\begin{array}{c}\tilde{\Delta}_{\mathbf{Q}} \\
    \tilde{\Lambda}_{\mathbf{Q}}\end{array}\right),
\end{eqnarray}
where now
\begin{eqnarray}
    \tilde{\Gamma}(\mathbf{Q})&=&-\left(\begin{matrix}
    U+U(\tilde{\Pi}_2+\tilde{\Pi}_3)U&2\alpha(U+U\tilde{\Pi}_2U)
    \\
    2\alpha(U+U\tilde{\Pi}_2U)& 4\alpha^2(U\tilde{\Pi}_2U)\end{matrix}\right).
    \label{eq:kend}
    \nonumber\\
\end{eqnarray}
In these equations, 
\begin{eqnarray}
    \tilde{\Delta}_{\mathbf{Q}}&=&\frac{1}{V}\sum_{\mathbf{k}}\langle \phi_{\downarrow}(-\mathbf{k}+\frac{\mathbf{Q}}{2})\phi_{\uparrow}(\mathbf{k}+\frac{\mathbf{Q}}{2})\rangle,
    \nonumber\\
    \tilde{\Lambda}_{\mathbf{Q}}&=&\frac{i}{V}\sum_{\mathbf{k}}(\mathbf{k}\times \mathbf{Q})
    \langle \phi_{\downarrow}(-\mathbf{k}+\frac{\mathbf{Q}}{2})\phi_{\uparrow}(\mathbf{k}+\frac{\mathbf{Q}}{2})\rangle,
\end{eqnarray}
and
\begin{eqnarray} \label{eq:PDW_Pi2_3}
    \tilde{\Pi}_2(\mathbf{Q})&=&\frac{1}{V\beta}\sum_{\mathbf{k},\omega_n}G_0(\mathbf{k}+\mathbf{Q},i\omega_n)G_0(\mathbf{k},-i\omega_n),
    \nonumber
    \\
    \tilde{\Pi}_3(\mathbf{Q})&=&\frac{4\alpha^2}{V\beta}\sum_{\mathbf{k},\omega_n}(\mathbf{k}\times\mathbf{Q})^2
    G_0(\mathbf{k}+\mathbf{Q},i\omega_n)G_0(\mathbf{k},-i\omega_n).
    \nonumber\\
\end{eqnarray}
The integrals in Eq.~(\ref{eq:PDW_Pi2_3}) are divergent in 2D. To handle such divergences 
we calculate them
in $d=2-2\epsilon$ dimensions. For $\vert \mathbf{Q}\vert <2k_F$, setting the mass $m=1$, a direct computation yields~\cite{serene1989stability,peskin2018introduction,nozieres1985bose}
\begin{eqnarray} \label{eq:minimal_subtraction}
    \tilde{\Pi}_2(\mathbf{Q})&=&\frac{1}{4\pi}\big[\epsilon^{-1}+\ln4\pi-\gamma-\ln(2\epsilon_{Q/2})\big]\label{eq:Pi2}
    \\
    \tilde{\Pi}_3(\mathbf{Q})&=&-4\alpha^2\frac{Q^2}{4\pi}
    \big\{\mu+(\epsilon_{q/2}-\mu)
    \nonumber
    \\
    &&\times\big[\epsilon^{-1}+\ln4\pi-\gamma-\ln(2\epsilon_{Q/2})\big] \big\}
    \label{eq:Pi3}
\end{eqnarray}
where $\gamma\simeq 0.5772$ is the Euler-Mascheroni constant. The divergences in Eq.~(\ref{eq:Pi2}) and Eq.~(\ref{eq:Pi3}) can in principle be removed by renormalizing the interaction $U$, the PSOI coupling $\alpha$, and the chemical potential $\mu$~\cite{andersen2004theory}. In this paper, for simplicity, we adopt the modified minimal subtraction scheme~\cite{peskin2018introduction}, which yields
\begin{eqnarray} \label{eq:minimal_subtraction}
\tilde{\Pi}_2^R(\mathbf{Q})&=&\frac{1}{4\pi}\ln\left(\frac{E_b}{2\epsilon_{Q/2}}\right),\label{eq:PiR2}
\\
\tilde{\Pi}_3^R(\mathbf{Q})&=&-4\alpha^2\frac{q^2}{4\pi}\left[\mu+(\epsilon_{Q/2}-\mu)\ln\left(\frac{E_b}{2\epsilon_{q/2}}\right)\right],\label{eq:PiR3}
\end{eqnarray}
where $E_b=(4\pi)^2U^{-2}$ is the two-particle binding energy~\cite{schmitt1989pairing,marsiglio2015pairing}. By substituting Eq.~(\ref{eq:PiR2}) and Eq.~(\ref{eq:PiR3}) into Eq.~(\ref{eq:kend}), we find that, when the parameter $r_s$ falls below a critical value, one eigenvalue of $\tilde{\Gamma}(\mathbf{Q})$ vanishes. 
It is important to note that this conclusion is independent of the direction of $\mathbf{Q}$.  
Since $\mathbf{Q}$ can be chosen in an arbitrary direction, this in turn implies a degeneracy of the energy of ground states featuring unidirectional mean-field
modulations~\cite{soto2014pair}.
Our mean-field results for many-body systems 
are in qualitative agreement with previous calculations for two-body PSOI problems~\cite{gindikin2018spin}.
The vanishing of one eigenvalue of $\tilde{\Gamma}(\mathbf{Q})$ at finite $\vert\mathbf{Q}\vert$ essentially mirrors the findings of~\cite{gindikin2018spin},
which concluded that pairs exhibit a moat-band dispersion with a minimum at finite relative momentum.
The critical value of $r_s$ and corresponding $q_c$, and their relation to the parameters $\alpha$ and $U$, are illustrated in Fig.~(\ref{fig:br1}).

From this result, we see that PSOI can induce not just the CDW and SCVW phases but also superconducting PDW phase. However, due to the subtraction of divergences performed in Eq.~(\ref{eq:minimal_subtraction}) it is not possible to directly compare the critical values of parameters at which CDW/SCVW and superconducting phase transitions occur. In practical situations, determining which phase transition occurs first requires further analysis, {\it i.e.} the implementation of a full renormalization scheme, which is beyond the scope of this paper.

\section{conclusion}\label{Con}
In this paper we have derived the Ginzburg-Landau effective theory for a PSOI two-dimensional electron system by using a double H-S transformation. The effective theory is written in terms of macroscopic averages of observable quantities, in particular the density and the vorticity of the spin current. These emerge naturally from the H-S decoupling of the Coulomb interaction and PSOI. 
By working at the level of Gaussian (mean-field) approximation, we find that the order parameter is a combination of CDW and spin-current-vorticity wave. The divergence of the correlation function for this order parameter reflects in the divergences of correlation functions of involved observables. In particular, of the density-density correlation function, thus explaining the observations of Ref.~\cite{gindikin2022electron}

On the other hand, because of the attractive interaction introduced by the PSOI term between pairs of electrons with certain spin and momentum configurations, we have investigated the possibility of superconducting broken phases characterised by the presence of a PDW state. We have found that the order parameter can acquire, among others, the form of an Amperean-like coupling. However, divergences that exist in the pair correlation functions, which require treatment with renormalization schemes, prevent us from directly comparing the PDW phase transition point with the non-superconducting ordered phase transitions discussed above. Further work, beyond the scope of the current paper, is needed to explore the relationship and competition between them. 


\begin{acknowledgments}
We acknowledge support from the European Commission under the EU Horizon 2020 MSCA-RISE-2019 programme (project 873028 HYDROTRONICS) and from the Leverhulme Trust under the grant agreement RPG-2019-363.     
\end{acknowledgments}



\normalem
\bibliography{v1}

\end{document}